\documentstyle[12pt]{article}
\newcounter{eqnn} \newcounter{eqs} \newcounter{secn}
\newcounter{subn}

\def\sec{\addtocounter{secn}{1}\setcounter{subn}{0}
\setcounter{eqs}{0}$\bf \thesecn $ }

\def\eqn{\addtocounter{eqs}{1}\;\;\;(\thesecn.\theeqs)}

\begin{document}

\begin{titlepage}

\pagestyle{empty} \begin{center} {\Large \bf         
}\end{center}              \begin{flushright}
LANCASTER-TH/9616\\              
September 1996\end{flushright}              \vfill              
\begin{center} {\LARGE Nucleosynthesis Bounds on Small Dirac Neutrino Masses 
due to Chiral Symmetry Breaking }\\ \end{center} \vfill \begin{center} 
{\bf John McDonald}\\              \vspace {0.1in}              
School of Physics and Chemistry, \\  Lancaster University,
\\  Lancaster, \\ LA1 4YB\\  United Kingdom 
\begin{footnote}{e-mail: jmcd@laxa.lancs.ac.uk}\end{footnote}
\begin{footnote}{Address from 1st October 1996: Dept. of Physics, 
P.O.Box 9, University of
Helsinki, FIN-00014 Helsinki, Finland}\end{footnote}
\\  \vfill              
\end{center}              \newpage \begin{center} {\bf
Abstract} \end{center}  

                   A Higgs doublet which has a positive mass 
squared term and Yukawa couplings to the quarks will acquire 
a vacuum expectation value typically of the order of $\rm
10^{2}eV$ or less, as a result of chiral symmetry breaking.
We consider nucleosynthesis constraints on models 
which use this fact as a basis for
understanding very small Dirac neutrino masses without
requiring very small Yukawa couplings. The simplest such 
model requires the introduction of a second Higgs
doublet plus right-handed neutrinos together with a global
horizontal symmetry $\rm U(1)_{H}$ which ensures that the 
neutrinos are naturally massless in the chiral symmetric limit. 
We show that present big-bang nucleosynthesis constraints impose 
a well-defined upper bound of 0.1eV on the neutrino
masses. This bound may become several orders of magnitude more
stringent in the future as our understanding of the
observational constraints on nucleosynthesis improves. 
 We discuss the phenomenological implications for neutrino dark matter, 
the solar neutrino problem and the atmospheric neutrino deficit.

\end{titlepage}
{\bf \sec. Introduction}

           The question of whether neutrinos have mass has
become particularly important in the light of evidence from
various experiments of a deficit in the number of electron
neutrinos from the Sun \cite{a1,a2,a3,a4,a5}, which cannot
be explained by modification of the standard solar model
\cite{a6} but which can readily be explained by neutrino
oscillations between massive neutrino species \cite{a7,a8}.
However, these experiments and others designed to directly
observe neutrino masses also require that neutrino masses
are much smaller than those of the charged leptons; in
particular, the electron neutrino mass must be less than 5eV, 
about $\rm 10^{-5}$ times the mass of the
electron \cite{a9}.

           In order for the $\rm
SU(3)_{c}xSU(2)_{L}xU(1)_{Y}$ Standard
Model to be able to account for neutrino masses, it must be 
extended by the addition of new particles. The simplest and
most commonly considered possibility is the addition of
right-handed weak isosinglet neutrinos. There are then
essentially two ways in which the small neutrino masses can
be understood. The simplest possibility is to couple
the right-handed neutrinos to the Higgs doublet via
extremely small Yukawa couplings, of magnitude $\rm
O(10^{-11})$ or less for the case of the electron neutrino.  
This will give rise to small Dirac masses for the neutrinos, 
with no lepton number (L) violation. The second possibility
is to introduce large L violating Majorana masses for the
right-handed neutrinos, together with Yukawa couplings of
the right-handed neutrinos to the Higgs doublet of a
strength unsuppressed relative to that of the charged lepton Yukawa
couplings. This results in the so-called "see-saw" mechanism 
for small neutrino masses \cite{a10}, in which mixing
between heavy right-handed neutrinos and left-handed
neutrinos results in essentially left-handed neutrinos with
a small, lepton-number violating Majorana mass. These are
the only possibilities for the case of the right-handed
neutrino extension of the Standard Model with only one
scalar doublet. However, if we consider models 
in which there is a second scalar doublet, then there is
another possibility. The right-handed neutrinos might couple 
only to the second scalar doublet, which might in turn
acquire only a very small vacuum expectation value (VEV).
The question of the smallness of the neutrino masses is then 
related to the reason for the smallness of the second Higgs
doublet's VEV. 

        One interesting possibility, originally suggested by Thomas and Xu
\cite{tx}, is that this small VEV could be induced by chiral symmetry breaking. 
They suggested a two 
Higgs doublet model with a global horizontal symmetry in
order to ensure the masslessness of the neutrinos in the
chiral symmetric limit. An upper bound was put on the
resulting Dirac neutrino masses by using limits on the rate
of helicity flipping processes in the supernova SN 1987A
\cite{tx,raffelt,burrows}. In the present paper, we will
reconsider this two Higgs doublet model. 
In particular, we will consider 
the constraints on such models coming from big-bang nucleosynthesis
\cite{bbn,kolb,olive2}. We will show that the present 
nucleosynthesis upper bound on the neutrino masses is of the same order
of magnitude as that coming from SN 1987A, but is much more
clearly defined than the supernova upper bound, which is
difficult to state precisely because of the complexity of
the physics of supernovae. Such a clearly defined upper
bound is important in order to be able to unambiguously
assess the implications of the model for neutrino
phenomenology, which is often sensitive to the 
mass squared of the neutrinos. In addition, the nucleosynthesis
 upper bound has the
possibility of becoming much tighter in the future as our
understanding of the observational constraints on big-bang
nucleosynthesis improves. The supernova upper bound, on the
other hand, is unlikely to be improved by much more than an
order of magnitude. 

              The paper is organized as follows. In section
2 we discuss the minimal two Higgs doublet model with a
global horizonal symmetry which can $naturally$ generate a
small Dirac neutrino mass via chiral symmetry breaking, as originally 
suggested by Thomas and Xu \cite{tx}.
 We also suggest a four Higgs doublet extension of 
the model which has the advantage of being compatible with
supersymmetry whilst not introducing large flavour
changing neutral current effects. In section 3 we consider the big
bang nucleosynthesis constraints on the resulting neutrino
masses. In section 4 we will consider the implications of
these constraints for various aspects of neutrino
phenomenology. In section 5 we give our conclusions.
\newpage{\bf \sec. Models for Dirac Neutrino Masses Due To
Chiral Symmetry Breaking} 

          We first consider the "minimal" model for Dirac
neutrino masses from chiral symmetry breaking, which was
first discussed in reference \cite{tx}. Consider the following two
Higgs doublet extension of the Standard Model. We will
refer to both scalar doublets as Higgs doublets, since they
will both acquire vacuum expectation values; however, only
one of them will have a negative mass squared term. Let
$\rm H_{+} = \left(\begin{array}{c} \phi_{+}^{+} \\
\phi_{+}^{0} \end{array} \right)$ be the second Higgs
doublet, assumed to have a positive mass squared, and let
$\rm H = \left( \begin{array}{c} \phi_{o}^{+} \\
\phi_{o}^{0} \end{array} \right)$ be the conventional
Standard Model Higgs doublet responsible for the quark and
lepton masses. In order that the neutrino masses can be
induced by chiral symmetry breaking we must impose on the
model that H does not couple to the right-handed neutrinos
and in addition that there are no scalar couplings of H to
$\rm H_{+}$ of the form $\rm H_{+}^{\dagger}H$, which would
otherwise lead to a large VEV for $\rm H_{+}$. We will
discuss shortly the necessary form of symmetry required to
achieve this. The most general form of Yukawa couplings is
then given by $${\rm h_{u}\overline{u}_{R}HQ +
h_{d}\overline{d}_{R} \tilde{H}Q +
h_{e}\overline{e}_{R}\tilde{H}L }$$ $${\rm
+\lambda_{\nu}\overline{\nu}_{R}H_{+}L +  
\lambda_{u}\overline{u}_{R}H_{+}Q +
\lambda_{d}\overline{d}_{R}\tilde{H}_{+}Q +
\lambda_{e}\overline{e}_{R}\tilde{H}_{+}L + h.c. \eqn},$$  
where $\rm \tilde{H}_{i} = \epsilon_{ij}H^{*}_{j}$, with i
and j being $\rm SU(2)_{L}$ indices and where we have
suppressed the generation indices. We will consider all
couplings to be real in the following. Once chiral symmetry
breaking occurs, the light quark condensates $\rm
<\overline{u}u>$, $\rm <\overline{d}d>$ and $\rm
<\overline{s}s>$ will become non-zero, with a value given by 
\cite{xsb}  $${\rm <\overline{q}q> =
\frac{f_{\pi}^{2}m_{\pi}^{2}} {\sqrt{2}(m_{u}+m_{d})}  
\eqn},$$ where $\rm q =$ u, d or s, $\rm f_{\pi}$ is the
pion decay constant, $\rm m_{\pi}$ is the pion mass and $\rm 
m_{u,d}$ are the current masses of the up and down quarks.
As a result, the leading terms in the scalar potential for
$\rm H_{+}$ will be of the form \cite{witten} $${\rm
V(H_{+}) = m_{+}^{2}|\phi_{+}^{0}|^{2}
-(\tilde{\lambda}_{q}<\overline{q}_{R}q_{L}>\phi_{+}^{0}
+\;\;h.c.) +\;\;... \eqn}.$$ (Throughout this paper we will
denote the Yukawa couplings in the mass eigenstate basis by
a tilde. By a choice of basis the neutrino Yukawa coupling
matrix $\rm \lambda_{\nu}$ can be made diagonal throughout). 
Thus the additional Higgs will gain a VEV $${\rm
<\phi_{+}^{0}>
=\frac{\tilde{\lambda}_{q}<\overline{q}q>}{m_{+}^{2}}
\eqn}.$$ With $\rm f_{\pi} \approx 90 MeV$, $\rm m_{u} +
m_{d} \approx 15MeV$ and $\rm m_{\pi} = 135MeV$ \cite{a9}
this gives  $${\rm <\phi_{+}^{0}> \approx
700\tilde{\lambda}_{q}
\left(\frac{100GeV}{m_{+}}\right)^{2}eV  \eqn},$$ and so the 
neutrinos will gain a mass given by  $${\rm m_{\nu} \approx
700 \lambda_{\nu} \tilde{\lambda}_{q}
\left(\frac{100GeV}{m_{+}}\right)^{2}eV   \eqn}.$$ We see
that with Yukawa couplings of magnitude $\rm ^{<}_{\sim} \;
0.1$ we would very naturally obtain neutrino masses of the
order of 1eV or less. 

                However, the above discussion does not
address the question of the conditions under which the
neutrinos remain massless in the chiral symmetric limit. The 
clearest approach is to impose a symmetry which can prevent
the coupling of the conventional Higgs to the right-handed
neutrinos or to $\rm H_{+}$, whilst allowing $\rm H_{+}$ to
couple to $\rm \overline{u}_{R}Q$ in order that chiral
symmetry breaking can induce a small VEV for $\rm H_{+}$.
Since the conventional Higgs must also couple to $\rm
\overline{u}_{R}Q$, the only possible way to have a such a
symmetry is to consider a symmetry that distinguishes
between different quark generations. The simplest
possibility is to consider a global horizontal symmetry $\rm 
U(1)_{H}$ under which $\rm H_{+}$, the first generation
right-handed up-type quark, $\rm u_{R_{1}}$, and the
right-handed neutrinos, $\rm \nu_{R_{i}}$, transform
according to $\rm (H_{+},u_{R_{1}},\nu_{R_{i}}) \rightarrow
e^{i\eta}(H_{+},u_{R_{1}},\nu_{R_{i}})$, with all other
fields invariant. In this case the allowed Yukawa couplings
from (2.1) are given by$${\rm h_{u\;\alpha
j}\overline{u}_{R\;\alpha}HQ_{j} +
h_{d\;ij}\overline{d}_{R\;i} \tilde{H}Q_{j} +
h_{e\;ij}\overline{e}_{R\;i}\tilde{H}L_{j} }$$ $${\rm
+\lambda_{\nu\;ij}\overline{\nu}_{R\;i}H_{+}L_{j} +  
\lambda_{u\;i}\overline{u}_{R\;1}H_{+}Q_{i} +  h.c.\eqn},$$
where $\rm i,\;j = 1,2,3$ and $\rm \alpha = 2,3$. In this
case the Dirac neutrino masses originate from the up quark
condensate. This model was originally suggested by
Thomas and Xu (equation (12) of reference \cite{tx}).
 The most important feature of the Yukawa
 couplings in (2.7) is that, in addition to the
neutrinos, the lightest up-type quark will also be massless
in the limit of unbroken chiral symmetry. This feature is an 
unavoidable consequence of ensuring the masslessness of the
neutrinos in the chiral symmetric limit via a symmetry.
However, as far as is known from the present understanding
of non-perturbative effects in QCD, such a possibility is
not inconsistent with hadron phenomenology
\cite{kaplan,banks}. It is possible that an up quark current 
mass could be generated by QCD instanton effects. (A massless 
down quark, on the other hand, would be inconsistent with
the observed pseudoscalar meson masses \cite{kaplan}).

         The form of the Yukawa couplings in (2.7) has an
important phenomenological advantage. In general one would
expect that adding a second Higgs doublet could result in
large flavour changing neutral current (FCNC) effects due to 
tree-level exchange of the additional neutral Higgs scalar
in $\rm H_{+}$ \cite{glashow,sher}. In particular, one would 
expect strong constraints to be imposed on $\rm \lambda_{q}$ 
by limits from $\rm \Delta m_{K}$ and $\rm \Delta m_{D}$.
However, it is easy to see that with the
form of the Yukawa coupling matrix $\rm \lambda_{u}$ in
(2.7) such tree-level processes do not occur. This is
because only the first generation right-handed up quark
couples to $\rm \lambda_{u}$. Therefore we only have a
coupling to $\rm \overline{u}_{R}c_{L}$ but not to $\rm
\overline{c}_{R}u_{L}$, as would be necessary in order to
have a tree-level contribution to $\rm \Delta m_{D}$. So
potentially dangerous tree-level FCNC effects are naturally
suppressed in this model.

         Before discussing the cosmological constraints,
 we briefly consider a supersymmetric version of the model. 
The minimal two Higgs doublet model given by equation (2.7) would not be
compatible with supersymmetry. A
supersymmetric version would require two more Higgs 
doublets; one in order to give masses to both the up-type and 
down-type quarks and one in order to maintain anomaly freedom
once the Higgsinos corresponding to $\rm H_{+}$ are
introduced \cite{nilles}. Thus a supersymmetric version of the model will have
four Higgs doublets: $\rm H_{u}$ and $\rm H_{d}$,
responsible for the up and down-type quark masses, and $\rm
H_{+}$ and $\rm H_{-}$, where $\rm H_{-}$ has the negative
of the hypercharge of $\rm H_{+}$. Although it would be possible for the 
additional Higgs doublet $\rm H_{-}$ not to couple to the
 quarks and leptons, in the most general case we would
expect both $\rm H_{d}$ and $\rm H_{-}$ to couple to $\rm
\overline{e}_{R}L$ and $\rm \overline{d}_{R}Q$. In
this case, for reasonable values of the mass of the
additional Higgs doublet (say $\rm m_{+} \; ^{<}_{\sim} \;
1TeV$), there would be a danger of a large tree-level
contribution to $\rm \Delta m_{K}$ \cite{glashow,sher}. 
 In order to avoid this danger we will extend the $\rm U(1)_{H}$ 
symmetry such that $\rm H_{-}$,
$\rm d_{R_{1}}$ and $\rm e_{R_{i}}$ transform according to 
$\rm (H_{-},d_{R_{1}},e_{R_{i}}) \rightarrow
e^{i\eta^{'}}(H_{-},d_{R_{1}},e_{R_{i}})$. 
In this case the Yukawa 
couplings of the four Higgs doublet model would be given by
$${\rm \lambda_{\nu\;ij}\overline{\nu}_{R\;i}H_{+}L_{j} +
\lambda_{u\;i}\overline{u}_{R\;1} H_{+}Q_{i} + h_{u\;\alpha
j}\overline{u}_{R\;\alpha}H_{u}Q_{j}  }$$ $${\rm
+\lambda_{e\;ij}\overline{e}_{R\;i}H_{-}L_{j} +  
\lambda_{d\;i}\overline{d}_{R\;1}H_{-}Q_{i} + h_{d\;\alpha
j}\overline{d}_{R\;\alpha}H_{d}Q_{j} \eqn}.$$ 
We see that this model requires that $\rm H_{-}$ develops a large
VEV in order to give a mass to the charged leptons and to
the down quark. 
In order for $\rm U(1)_{H}$ to prevent $\rm H_{+}$ from
coupling to $\rm H_{-}$, it is necessary that the
charge of $\rm H_{-}$ under $\rm U(1)_{H}$ should not equal
the negative of the charge of $\rm H_{+}$ ($\rm \eta^{'} \neq -\eta$). 

           In order to discuss these models further,
 we must consider the observational
constraints on the product $\rm \lambda_{\nu}
\tilde{\lambda}_{q}$ appearing in the expression for the
neutrino masses (2.6). These are imposed by primordial
nucleosynthesis \cite{bbn,kolb,olive2} and by limits on
neutrino helicity-changing processes from the supernova SN
1987A \cite{tx,raffelt,burrows}. Since the neutrino masses
and the observational constraints in the case of the supersymmetrizible four
Higgs doublet model will be essentially the same as those in 
the case of the "minimal" two Higgs doublet model, we will
focus our attention on the two Higgs doublet model in the
following.   \newpage {\bf \sec. Nucleosynthesis Constraints 
on the Neutrino Masses}     

             The success of the primordial nucleosynthesis
calculation of the abundances of light elements (D,
$\rm^{3}He$, $\rm ^{4}He$ and $\rm ^{7}Li$) in the Standard
Model \cite{bbn,kolb} imposes strong constraints on the
addition of new light particles of mass less than O(1)MeV,
such as light right-handed neutrinos. There has recently
been some controversy over exactly what the nucleosynthesis
upper bound on the number of additional light degrees of
freedom is \cite{olive2}. We may discuss this in terms of
the effective number of massless left-handed neutrinos at
nucleosynthesis, $\rm N_{\nu}$. The controversy is related
to what observational constraints on light element
abundances should be imposed. Previously an indirect bound
on the primordial $\rm D + ^{3}He$ abundance was used,
inferred by using chemical evolution models combined with
measurements of the abundance in the solar neighbourhood.
This gave an upper bound $\rm N_{\nu} \; ^{<}_{\sim} \; 3.3$ 
\cite{bbn}. However, recent evidence from planetary nebulae
implies a need for $\rm ^{3}He$ production in low mass
stars, suggesting that the primordial $\rm D + ^{3}He$
density inferred from chemical evolution models is incorrect 
\cite{olive2}. 
Using instead the more reliable estimate of
the primordial $\rm ^{7}Li$ density, 
Olive et al give a $\rm 95 \%$c.l.
upper limit $\rm N_{\nu} < 3.9$ with a central value for
$\rm N_{\nu}$ equal to 3.02 \cite{olive2}. 
Kernan and Sarkar 
conclude that the upper bound can be as large as 4.53, 
taking observational uncertainties into account \cite{sarkar}.
The constraints
on the neutrino masses following from these upper bounds on
$\rm N_{\nu}$ will depend on how many neutrinos are
effectively massless at nucleosynthesis. Present
experimental constraints give $\rm m_{\nu_{e}} < 5.1eV$,
$\rm m_{\nu_{\mu}} < 160keV$ and $\rm m_{\nu_{\tau}} <
24MeV$ \cite{a9}. Thus it is possible that $\rm
m_{\nu_{\tau}}$ could be heavier than 1MeV and so not affect 
nucleosynthesis. However, bounds from the supernova SN 1987A 
combined with constraints from nucleosynthesis imply that
$\rm m_{\nu_{\tau}}$ must be less than 0.4MeV if the
dominant $\rm \nu_{\tau}$ decay is to electromagnetic final
states \cite{sigl}, as would be the case in the class of
model we are discussing here. Thus from the point of view of
nucleosynthesis we will consider all three neutrino species
be effectively massless.

       The condition $\rm \Delta N_{\nu} < 1.53$ (where we
define $\rm \Delta N_{\nu}$ by $\rm N_{\nu} = 3 + \Delta
N_{\nu}$) requires that the right-handed neutrinos freeze
out of chemical equilibrium prior to the quark-hadron phase
transition. This is because each right-handed neutrino
species contributes the equivalent of $\rm n_{eff}$ 
left-handed neutrino species to the energy density at
nucleosynthesis, where $\rm n_{eff}$ is related to the
freeze-out temperature $\rm T_{fr}$ of the right-handed
neutrinos by $${\rm n_{eff} = \left(
\frac{g(T_{nuc})}{g(T_{fr})} \right)^{4/3} \eqn},$$ where
$\rm g(T) = g_{b} + \frac{7}{8}g_{f}$ is the number of 
effectively massless degrees of freedom in thermal
equilibrium at temperature T, with $\rm g_{b} = 2$ for the
photon and $\rm g_{f} = 4$ for Dirac fermions \cite{kolb}.
The reduction of $\rm n_{eff}$ from 1 is due to the
adiabatic expansion of the Universe when the number of
effectively massless degrees of freedom changes, for example 
during a confining phase transition or when a particle
species becomes non-relativistic and annihilates away; this
will dilute a particle species which is out of chemical
equilibrium relative to those in equilibrium, which maintain 
their equilibrium densities. At temperatures above the
quark-hadron phase transition, one has free quarks and
gluons in thermal equilibrium, and $\rm g(T) = 61.75$ at
$\rm T \approx T_{qh}$. Below the temperature of the
quark-hadron phase transition, one has quarks and gluons
confined in hadrons, with $\rm g(T_{qh}) = 17.25$ and $\rm
g(T_{nuc}) = 10.75$. From this we see that for $\rm T_{fr}$
slightly below $\rm T_{qh}$ we have $\rm n_{eff} = 0.53$,
whilst for $\rm T_{fr}$ slightly above $\rm T_{qh}$ this
becomes $\rm n_{eff} = 0.097$. Thus for the case where we
have 3 effectively massless right-handed neutrinos we find
that for $\rm T_{fr}$ slightly above $\rm T_{qh}$ we have
$\rm \Delta N_{\nu} = 0.29$, whilst for $\rm T_{fr}$
slightly below $\rm T_{qh}$ we have $\rm \Delta N_{\nu} =
1.60$. Thus we see that $\rm T_{fr} < T_{qh}$ is ruled out.
In Table 1 we list the values of $\rm \Delta N_{\nu}$ as a
function of the known Standard Model particle thresholds.
(The inclusion of the Higgs boson thresholds due to the
physical Higgs of the Standard Model and the additional
doublet $\rm H_{+}$ in the two Higgs doublet model will only
reduce the smallest possible value of $\rm \Delta N_{\nu}$
from 0.14 to 0.13). 

      Imposing the condition $\rm T_{fr} > T_{qh}$ allows us 
to put an upper bound on the couplings entering in the
cross-sections for processes changing the number of
right-handed neutrinos via $\rm \phi_{+}^{0}$ exchange: (i)
$\rm \overline{\nu}_{R} \nu_{R} \leftrightarrow
\overline{\nu}_{L} \nu_{L}$ (Figure 1) and (ii) $\rm
\overline{\nu}_{R} \nu_{L} \leftrightarrow  
\overline{q}_{R} q_{L}$ and related inelastic scattering
processes (Figure 2). In addition there are analogous
processes formed by replacing $\rm \phi_{+}^{0}$ by $\rm
\phi_{+}^{+}$. Adding
these simply multiplies the rate of $\rm \nu_{R\;i}$
annihilation due to $\rm \phi_{+}^{0}$ by a factor of 2.
From the diagram of Figure 1 we obtain for the $\rm
\phi_{+}^{0}$ contribution to the $\rm \nu_{R\;i}$
annihilation cross-section $${\rm
\sigma(\overline{\nu}_{R\;i} \nu_{R\;i} \rightarrow  
\overline{\nu}_{L} \nu_{L}) = \frac{E^{2}}{24\pi m_{+}^{4}}
\sum_{j} \left(\lambda_{\nu\;i} \lambda_{\nu\;j}\right)^{2}  
\eqn}$$ where E is the energy of each annihilating neutrino
in the centre of mass frame. In this we have assumed that E
is small compared with $\rm \frac{m_{+}}{2}$. If E were to
approach or exceed $\rm \frac{m_{+}}{2}$ then the effect of
the s-channel pole terms or of the direct production of $\rm 
H_{+}$ Higgs scalars by neutrino annihilations would simply 
be to tighten the upper bounds we derive below. Including
the $\rm \phi_{+}^{+}$ exchange contribution, the
annihilation rate in the early Universe is then given by
$${\rm \Gamma_{ann} = 2 <n\sigma v> \approx \frac{9 T^{5}}{8
\pi^{2} m_{+}^{4}} \sum_{j} \left(\lambda_{\nu \; i}
\lambda_{\nu \; j}\right)^{2}  \eqn},$$ where n is the
number density of scattering particles in the thermal
background ($\rm n = \frac{1.2 g^{`}}{\pi^{2}}T^{3}$, with
$\rm g^{`} = g_{B} + \frac{3}{4}g_{F}$ being the number of
light degrees of freedom in thermal equilibrium), v is the
relative velocity of the scattering particles ($\rm v=1$)
and we have used for the average energy of the annihilating
fermions $\rm E \approx 3T$ \cite{kolb}. Requiring that this
is less than the expansion rate H of the Universe at the
temperature of the quark-hadron phase transition $\rm
T_{qh}$ ($\rm H = \frac{k_{T}T_{qh}^{2}}{M_{Pl}}$, where
$\rm k_{T} = \left(\frac{4 \pi^{3}
g(T_{qh})}{45}\right)^{1/2} \approx 13$) then gives the
upper bound $${\rm \left( \sum_{j} ( \lambda_{\nu \; i}
\lambda_{\nu \; j})^{2} \right)^{1/4} \;\; ^{<}_{\sim} \;\;
0.024 \left(\frac{m_{+}}{100GeV}\right) 
\left(\frac{0.2GeV}{T_{qh}}\right)^{3/4} 
\left(\frac{g(T_{fr})}{g(T_{qh})}\right)^{1/8}
\left(\frac{T_{qh}}{T_{fr}}\right)^{3/4} \eqn},$$ where we
have used $\rm T_{qh} = 200MeV$ as a typical value \cite{qh} 
and we have shown explicitly the dependence on $\rm T_{fr}$. 
(Note that, for the case $\rm i = j $, this gives an upper
bound on $\rm \lambda_{\nu \;i}$ itself). We can also obtain 
an upper bound on the product $\rm \lambda_{\nu \;
i}\tilde{\lambda}_{u}$ which enters in the expression for
the neutrino mass. Adding the rates for the processes shown
in Figure 2, (i) $\rm \overline{\nu}_{R\;i} \nu_{L\;i}
\rightarrow \overline{q}_{L} q_{R}$, (ii) $\rm
\overline{\nu}_{R\;i} q_{R} \rightarrow
\overline{\nu}_{L\;i} q_{L}$ and (iii) $\rm
\overline{\nu}_{R\;i} \overline{q}_{L} \rightarrow
\overline{\nu}_{L\;i} \overline{q}_{R}$, and including the
factor of 2 for the analogous $\rm \phi_{+}^{+}$ exchange
processes, we obtain  $${\rm \Gamma \approx \frac{135
T^{5}}{8 \pi^{3} m_{+}^{4}} \lambda_{\nu \;i }^{\dagger}
\lambda_{\nu\;i} Tr[\lambda_{u}^{\dagger} \lambda_{u}]   
\eqn},$$ where the trace is over quark flavours. Requiring
that this be less than the expansion rate of the Universe at 
the quark-hadron phase transition then gives the upper bound  
$${\rm (\lambda_{\nu \; i}^{\dagger}\lambda_{\nu \;
i})^{1/2} (Tr[\lambda_{u}^{\dagger}\lambda_{u}])^{1/2} \;
^{<}_{\sim} \; 1.6x10^{-4}
\left(\frac{m_{+}}{100GeV}\right)^{2}
\left(\frac{0.2GeV}{T_{qh}}\right)^{3/2}
\left(\frac{g(T_{fr})}{g(T_{qh})}\right)^{1/4}
\left(\frac{T_{qh}}{T_{fr}}\right)^{3/2}  \eqn}.$$ Note
that, when combined with the upper bound on $\rm
\lambda_{\nu \; i}$ from (3.4), this gives an upper bound on 
$\rm Tr[\lambda_{u}^{\dagger}\lambda_{u}]$ itself. Using
(3.6), and noting that $\rm \tilde{\lambda}_{u} \; \leq
Tr[\lambda_{u}^{\dagger} \lambda_{u}]^{1/2}$, we see that
the upper limit on the neutrino masses is then given by  
$${\rm m_{\nu} \; ^{<}_{\sim} \; 0.11eV
\left(\frac{0.2GeV}{T_{qh}}\right)^{3/2}  
\left(\frac{g(T_{fr})}{g(T_{qh})}\right)^{1/4}
\left(\frac{T_{qh}}{T_{fr}}\right)^{3/2} \eqn}.$$ Thus we
see that the present nucleosynthesis constraint $\rm \Delta
N_{\nu} < 0.9$ implies that the neutrino masses must be less 
that about 0.1eV in this class of model. The nucleosynthesis 
upper bound on the neutrino masses could become much more
stringent in the future if the upper bound on $\rm N_{\nu}$
were to approach the Standard Model value $\rm N_{\nu} = 3$. 
From Table 1 we see that once the limit on $\rm \Delta
N_{\nu}$ is less than 0.29, the freeze-out temperature must
be larger than $\rm \frac{m_{c}}{3}$ (corresponding to $\rm
E_{\nu} \approx 3T > m_{c}$), leading to an upper bound on
the neutrino masses of $\rm 2.6x10^{-2}eV$, whilst if the
limit on $\rm \Delta N_{\nu}$ were to become less than 0.20, 
the freeze-out temperature would have to be larger than $\rm
\frac{m_{W}}{3}$, giving an upper bound on the neutrino
masses of about $\rm 6.3x10^{-5}eV$. If $\rm \Delta N_{\nu}$ 
was constrained to be below 0.14 (or 0.13, including Higgs
thresholds in the two Higgs doublet model), then it would
not be possible for three massless right-handed neutrinos to 
be consistent with nucleosynthesis when only the thresholds
of the Standard Model or its two Higgs doublet extension are 
considered. The four Higgs doublet supersymmetric extension
could, however, allow for a significantly smaller $\rm
\Delta N_{\nu}$ if the freeze-out temperature was
larger than the masses of the supersymmetric partners of the 
Standard Model particles.

       We next compare the upper bound on the neutrino
masses coming from nucleosynthesis with that coming from
neutrino helicity-flip processes in the supernova SN 1987A
\cite{tx,raffelt,burrows}. Thomas and Xu give an upper bound 
on the neutrino masses of 0.05eV \cite{tx}, based on an
upper bound on the helicity flipping cross-section from the 
requirement that the energy of the supernova is not rapidly
carried away by right-handed neutrino emission, which would
unacceptably shorten the observed neutrino pulse. However,
it is difficult to give anything better than an order of
magnitude estimate for the neutrino mass upper bound from
the supernova \cite{raffelt}. The physics of the interaction 
of neutrinos with nucleons in the dense core of the
supernova is a non-trivial problem in nuclear physics, which 
introduces an order of magnitude uncertainty in the upper
bound on $\rm (\lambda_{\nu}\tilde{\lambda}_{q})^{2}$
\cite{raffelt}. The bound derived by Thomas and Xu is based
on evaluating $\rm <n\;| \overline{u}u|\;n>$ for a single
isolated nucleon \cite{tx}, whereas in the dense core of the 
supernova collective effects might increase this scattering
rate by perhaps an order of magnitude \cite{raffelt}, in
which case the upper bound on the neutrino masses from the
supernova becomes $\rm 0.16eV$. Another order of magnitude
uncertainty in the right-handed neutrino emission rate is
introduced by the model of the supernova itself
\cite{raffelt}. Thus we see that it is difficult to be very
clear about exactly what the upper limit from the supernova
is, beyond giving an order of magnitude estimate which is
typically somewhere between 0.1eV and 0.01eV. In contrast
with the supernova bound, the nucleosynthesis upper bound on 
the neutrino masses is much simpler to derive and is quite
precise. Since the phenomenology of neutrinos is often
sensitive to the neutrino mass squared, such precision in
the neutrino mass bounds is important in order to be able to 
unambiguously discuss their phenomenological implications. 
The nucleosynthesis upper bound also has the possibility of
becoming several orders of magnitude tighter in the future
as our understanding of big-bang nucleosynthesis develops,
whereas the supernova upper bound is unlikely to be improved 
by much more than an order of magnitude \cite{raffelt}.
\newpage{\bf \sec. Aspects of the phenomenology of massive
neutrinos with masses induced by chiral symmetry breaking}.

        We next compare the allowed range of neutrino masses 
from chiral symmetry breaking with the range of neutrino
masses required in order to explain certain important
observations in cosmology and astrophysics, namely a) the
possibility of a hot component of dark matter due to a
massive neutrino \cite{nudm} b) the solar neutrino problem 
\cite{a7,a8} and c) the atmospheric neutrino deficit
\cite{atmnu}. \newline \underline{a) Hot dark matter}: We
can immediately see that there is no possibility of a
significant contribution to cosmological dark matter from
neutrinos in this model. This would require a neutrino mass
in the 1eV to 10eV range \cite{kolb,nudm}, whereas the
present nucleosynthesis constraint gives an upper bound of
about 0.11eV. This would allow a fractional contribution to
the closure dark matter density of no more than $\rm
\Omega_{\nu} \approx 0.053 \left(\frac{m_{\nu}}{5eV}\right)
h^{-2} \; ^{<}_{\sim} \; 0.005$, where h is the Hubble
parameter in units of $\rm 100 \; km \; s^{-1} \; Mpc^{-1}$
($\rm 0.5 \; ^{<}_{\sim} \; h \; ^{<}_{\sim} \; 1$)
\cite{kolb}. Thus this mechanism for neutrino masses is not
compatible with recent ideas which use a hot neutrino
component of dark matter  (with $\rm \Omega_{\nu} \approx
0.2$) to account for the discrepencies between observed
large-scale structure and the large-scale structure
predicted by $\rm \Omega = 1$ cold dark matter models with a
scale-invariant primordial fluctuation spectrum
\cite{nudm}.  \newline \underline{b) MSW solution of the
solar neutrino problem}: The Mikheyev-Smirnov-Wolfenstein
(MSW) matter oscillation solution to the solar neutrino
problem \cite{a7,a8} has three main regions of mass squared
splitting between the neutrino masses; a solution
corresponding to $\rm \Delta m_{\nu}^{2} \approx
10^{-4}eV^{2}$ for neutrino mixing angles $\rm \theta$
corresponding to $\rm Sin^{2}2\theta$ increasing from $\rm
10^{-4}$ to $\rm 10^{-1}$ (adiabatic solution), a second
solution corresponding to $\rm \Delta m_{\nu}^{2}$
decreasing from $\rm 10^{-4}eV^{2}$ to $\rm 10^{-7}eV^{2}$
as $\rm Sin^{2}2\theta$ increases from $\rm 10^{-4}$ to $\rm 
10^{-1}$ (non-adiabatic solution) and a third region
corresponding to $\rm Sin^{2}2\theta$ of order 1 for mass
squared splittings from about $\rm 10^{-7}eV^{2}$ up to
about $\rm 10^{-4}eV^{2}$ (large-mixing solution).  Thus the 
MSW solution is generally consistent with the present
nucleosynthesis upper bound on the neutrino masses. However, 
from Table 1 we see that if the nucleosynthesis constraint
on $\rm \Delta N_{\nu}$ should become established at $\rm
\Delta N_{\nu} < 0.24$, then the upper limit on the neutrino 
masses would become $\rm m_{\nu} < 4.1x10^{-3}eV$, which
would rule out the adiabatic solution, whilst if the upper
bound was tightened to $\rm \Delta N_{\nu} < 0.20$, then the 
upper bound would become $\rm m_{\nu} < 6.3x10^{-5}eV$ and
the MSW solution to the solar neutrino problem would be
ruled out completely.  \newline c) \underline{Atmospheric
neutrino deficit}: The atmospheric neutrino deficit problem
\cite{atmnu} refers to the observation that the electron and 
muon neutrinos arising from high energy cosmic rays incident 
on the upper atmosphere should be observed with the ratio
$\rm \frac{(\nu_{\mu} + \overline{\nu}_{\mu})}{(\nu_{e}+
\overline{\nu}_{e})} \approx 2$. However, experimentally
this ratio is observed to be approximately equal to 1, which 
may be interpreted as evidence of vacuum oscillations of 
muon neutrinos to $\rm \nu_{e}$ or $\rm \nu_{\tau}$ with a
large vacuum mixing angle ($\rm Sin^{2}2\theta \;
^{>}_{\sim} \; 0.4$) and $\rm \Delta m_{\nu}^{2} \approx
10^{-3}eV^{2}$ to $\rm 10^{-1}eV^{2}$ \cite{atmnu}. From
this we see that the range of mass squared splitting
required to solve the atmospheric neutrino problem is
consistent with the present nucleosynthesis constraint ($\rm 
\Delta m_{\nu}^{2} \; ^{<}_{\sim} \; 10^{-2}eV^{2}$).
However, should the nucleosynthesis constraint be improved
to $\rm \Delta N_{\nu} < 0.24$, then the upper bound on the
neutrino masses would become $\rm m_{\nu} < 4.1x10^{-3}eV$,
which would rule out a vacuum oscillation soultion to the
atmospheric neutrino deficit problem. 
\newpage {\bf \sec. Conclusions.}

            We have considered the possibility that small
Dirac neutrino masses in the Standard Model could arise as a 
result of a small expectation value induced in a second
Higgs field by chiral symmetry breaking. Present big bang
nucleosynthesis constraints imply that the neutrino masses
are less than about 0.11eV. This upper bound may become much
more stringent in the future, as our understanding of the
primordial light element abundances improves and the
constraint on the number of additional light degrees of
freedom compatible with big bang nucleosynthesis becomes
tighter. The constraints on the neutrino masses from the 
rate of helicity flip in the supernova SN 1987A are not
significantly stronger than the present nucleosynthesis
constraint and, unlike the nucleosynthesis constraint, are
difficult to state precisely and not likely to become much
tighter in the future.

        In general big bang nucleosynthesis rules out
neutrinos with masses due to chiral symmetry breaking as a
significant contributor to cosmological dark matter, but
does not at present significantly constrain the possibility
of an MSW solution of the solar neutrino problem or a vacuum
oscillation solution of the atmospheric neutrino deficit.
However, should the nucleosynthesis constraint on the number 
of additional neutrinos become established in the future at
$\rm \Delta N_{\nu} < 0.24$, then the vacuum oscillation
solution of the atmospheric neutrino deficit and the
adiabatic MSW solution of the solar neutrino problem would
be ruled out, whilst if the bound were to become established 
at $\rm \Delta N_{\nu} < 0.20$ the MSW solution of the solar 
neutrino problem would be ruled out completely. Thus a substantial
improvement in the present bound on $\rm \Delta N_{\nu}$ 
could impose severe constraints on
neutrino phenomenology in this class of neutrino mass model.  

      The author would like to thank Emilio Ribeiro and
Pedro Bicudo for useful discussions regarding chiral
symmetry breaking and Subir Sarkar 
for some informative comments.
This research was funded by the Grupo Teorico da Altas Energias
(GTAE), Portugal and by the PPARC, UK.    \newpage {\bf Figure 
Captions.} \newline \newline \newline Figure 1: Annihilation 
of $\rm \nu_{R} \overline{\nu}_{R} $ pairs. \newline
\newline Figure 2: Inelastic scattering processes involving
thermal background quarks which change the number of
right-handed neutrinos. \newline \newline \newpage {\bf
Table 1. $\rm \Delta N_{\nu}$ due to three right-handed
neutrinos and the upper bound on $\rm m_{\nu}$ as a function 
of $\rm T_{fr}$ } \newline \newline  \begin{table}[h]
\begin{center} $$\rm \begin{array}{ccc} \hline \hline
\multicolumn{1}{c}{\Delta N_{\nu}} &
\multicolumn{1}{c}{m_{\nu}(eV) <} &
\multicolumn{1}{c}{T_{fr} <} \\ \hline  3.00& \;\;\;Ruled \;
Out \;\;\;& m_{\mu}/3 \\ 2.06&  Ruled \; Out & m_{\pi}/3 \\
1.60& Ruled \; Out & T_{qh} \\ 0.29 & 0.11 & m_{c}/3 \\ 0.24 
& 2.6x10^{-2} & m_{b}/3 \\ 0.20 & 4.1x10^{-3} & m_{W}/3\\
0.17 & 6.3x10^{-5} & m_{Z}/3 \\ 0.16 & 5.2x10^{-5} & m_{t}/3 
\\ 0.14 & 2.0x10^{-5} & ? \\ \end{array}$$ \end{center}  
\end{table}

 \end{document}